\documentclass[11pt,a4paper]{article}
\usepackage{jcappub}
\usepackage[english]{babel}
\usepackage[utf8]{inputenc}
\usepackage{amsmath}
\usepackage{graphicx}
\usepackage[colorinlistoftodos]{todonotes}
\usepackage{color}
\usepackage{hyperref}
\usepackage[title,toc,titletoc]{appendix}

\begin{document}

\title{Inflationary Constraints on the Van der Waals Equation of State}

\author[a]{G. Vardiashvili,}
\author[a]{E. Halstead,}
\author[b]{R. Poltis,}
\author[a]{A. Morgan,}
\author[a]{and D. Tobar,}
\emailAdd{gvardias@skidmore.edu}
\emailAdd{ehalstea@skidmore.edu}
\emailAdd{rpoltis@niagara.edu}
\emailAdd{amorgan@skidmore.edu}
\emailAdd{dtobar@skidmore.edu}
\affiliation[a]{Physics Department, Skidmore College, Saratoga Springs, NY 12866}
\affiliation[b]{Department of Physics, Niagara University, Lewiston, NY 14109}

\date{\today}

\abstract{
In spite of the strong observational evidence suggesting a period of rapid expansion in the early universe, the identity of the inflaton field that drove this expansion remains elusive. Many inflaton candidate particles (both known and hypothesized) have been proposed to explain the early accelerated expansion of the universe. Other explanations for an era of rapid expansion in the early universe have been proposed via modifications of gravity in one form or another.

In this paper, we consider the possibility of using a Van der Waals equation of state to describe the early-time accelerated expansion of the universe. We do not attempt to explain {\it why} the early universe may be filled with such a fluid, but rather investigate what constraints may be placed on the parameters which describe a Van der Waals fluid that sources an accelerated expansion in the early universe. We consider three different versions of the Van der Waals equation of state and constrain the parameters of each using CMB data. We find that two of the models do not fit constraints from observation. A third model may satisfy observational constraints, but only under very narrow circumstances.}

\maketitle
\flushbottom

\section{Introduction}

The discovery of the late time accelerated expansion of the universe \cite{bahcall1999cosmic} motivated the search for physical theories that naturally contain a dark fluid with a nearly constant energy density and negative pressure. Perhaps the simplest cosmological model that incorporates this is the $\Lambda CDM$ model \cite{carroll1992cosmological}, which introduces a cosmological constant $\Lambda $ into the Einstein field equations to explain the late-time accelerated expansion. The many-orders of magnitude disconnect between the expected value of $\Lambda$ from particle physics and the value of $\Lambda$ inferred from cosmological observation motivates the search for models of dark energy beyond $\Lambda CDM$. The details of dark energy based cosmology can be seen in ref.~\cite{Frieman:2008sn}. Models that extend beyond $\Lambda CDM$ include scalar field inflationary models \cite{Starobinsky:1998fr}, and in recent years various models have been proposed with a modified equation of state \cite{caldwell1998cosmological, Cardone:2005ut}. Chaplygin gas \cite{Kamenshchik:2001cp, del2013single, Fabris:2001tm,Kremer:2003ev, Fabris:2010vd, Kumar:2014pia, Avelino:2014nva, Kahya:2015pza, Sharif:2014yga, Sharov:2015ifa, Gorini:2004by, Villanueva:2015ypa, Bento:2002yx} and Van der Waals quintessence \cite{capozziello2002van, kremer2003cosmological, Capozziello:2003tk, Kremer:2004di, Capozziello:2004ej, Pourhassan:2013wza, Kremer:2003vs} models, for example, provide accurate qualitative descriptions of the universal expansion in different epochs of the cosmic evolution. These models can be constrained not only from observations of the current epoch of a late-time accelerated expansion, but may also be cast in the form of a scalar field in the early universe. Consequentially, these models also make inflationary predictions that can be tested against observation \cite{del2013single}. 

The authors of ref.~\cite{capozziello2002van} propose a model of the universe that contains a baryonic component with a barotropic fluid equation of state ($p=w\rho$) and a dark fluid component characterized by a Van der Waals equation of state that causes a late-time accelerated expansion. Unlike conventional barotropic models, a Van der Waals fluid contains phase transitions between different cosmological eras. One intriguing aspect of the Van der Waals fluid model proposed in ref.~\cite{capozziello2002van} is that it also contains an early-time de-Sitter expansion followed by a matter-dominated epoch without the introduction of a separate scalar field. In other words, the authors find that both an early and late time accelerated expansion (e.g. inflation and dark energy) may be caused by the same Van der Waals fluid.

In this work, we explore three different parametrizations of the Van der Waals equation of state describing an early time accelerated expansion.  We choose a series of observational constraints based on data from the Planck satellite \cite{Ade:2015xua} and use these constraints to place limits on the parameters of each model. In section \ref{sec:TheVdWEoS} we discuss cosmology with a Van der Waals equation of state and introduce the three Van der Waals models that we consider. In section \ref{sec:InflationwithVdWFluid} we develop the formalism and list the criteria that we will use to evaluate each model. Sections \ref{sec:oneparametermodel}, \ref{sec:twoparametermodel}, and \ref{sec:threeparametermodel} discuss the results from each of the three models. Conclusions are given in section \ref{sec:Conclusions}.

\section{The Van der Waals equation of state}\label{sec:TheVdWEoS}
We begin with the homogenous and isotropic Friedmann-Robertson-Walker (FRW) metric
\begin{equation}
ds^2=dt^2-a(t)^2\left[dr^2-r^2d\Omega\right],
\end{equation}
from which one may derive the Friedmann equations (without a cosmological constant term) as
\begin{eqnarray}\label{Friedmann1}
\left(\frac{\dot{a}}{a} \right)^2+\frac{k}{a^2} = \frac{8 \pi G}{3} \rho \\
\label{Friedmann2}
\frac{\ddot{a}}{a} = -\frac{4 \pi G}{3}\left(\rho+3p\right)
\end{eqnarray}
 where $a$ is the scale factor, an overdot indicates differentiation with respect time, $k$ is the curvature parameter,  and $\rho$ and $p$ are the energy density and pressure of the cosmological fluid, respectively. The curvature parameter $k$ is set to zero, as current observations \cite{Ade:2015xua} indicate that the universe is spatially flat. 

Local conservation of stress-energy for the cosmological fluid yields
\begin{equation}
\label{rhodot}
\dot{\rho}=-3\frac{\dot{a}}{a}\left(\rho+p\right).
\end{equation}
In the scalar field description, the energy density and pressure of the scalar field are \cite{lidsey1997reconstructing}  
\begin{eqnarray}\label{inflrhoP}
\rho &=& \frac{1}{2}\dot{\phi}^2+V(\phi) \\
\label{inflP}
p &=& \frac{1}{2}\dot{\phi}^2-V(\phi).
\end{eqnarray} 
Typically, the equation of state of a cosmological fluid is described by the parameter $w=p/\rho=(\frac{1}{2}\dot{\phi}^2-V)/(\frac{1}{2}\dot{\phi}^2+V)$, where the second equality is for the specific case of a scalar field. This equation of state produces a nonlinear relationship between the pressure and energy density of the scalar field, whose dynamics are determined by the form of the potential $V(\phi)$.

We now consider three models in which the inflationary expansion of the early universe is posited to be driven by a Van der Waals fluid. Using a scalar field treatment, we extract constraints on the parameters in the Van der Waals equation of state from inflationary observables and a few reasonable assumptions (described in section \ref{sec:InflationwithVdWFluid} below). We would like to mention here that we are not proposing a mechanism that explains \textit{why} the inflationary fluid is described by this equation of state. Rather, given our current lack of understanding of the nature of inflation, this equation of state may be thought of as a generalization of the conventional equation of state $p=w\rho$. The three Van der Waals models that we consider are described below.

\begin{enumerate}
\item \textbf{One-parameter model:} The first model is a reduced, dimensionless equation of state given by
\begin{equation}
\label{eq:1parameterEoS}
p=\frac{8w\rho}{3-\rho}-3\rho^2,
\end{equation}
where $w$ is a non-negative constant.\cite{jantsch2016van, Kremer:2003vs}

\item
\textbf{Two-parameter model:} The second model parametrizes the Van der Waals equation of state in terms of a constant parameter $\gamma$ and critical point $\rho_c$ (see \cite{Capozziello:2003tk} for details). The equation of state for this two-parameter model is
\begin{equation}
\label{eq:2parameterEoS}
p=\frac{\gamma\rho}{1-\frac{1}{3}\frac{\rho}{\rho_c}}-\frac{9\gamma}{8\rho_c}\rho^2.
\end{equation}

\item
\textbf{Three-parameter model:} 
The third model that we consider is an equation of state based on three constant parameters $\alpha$, $\beta$, and $\gamma$. In a classical Van der Waals gas the $\alpha$ term is related to the intermolecular interaction and the $\beta$ term is related to the particle size. This model is described by the equation of state
\begin{equation}
\label{eq:3parameterEoS}
p=\frac{\gamma \rho}{1-\beta \rho}-\alpha \rho^2.
\end{equation}
\end{enumerate}

\section{Inflation with Van der Waals fluid}\label{sec:InflationwithVdWFluid}
Beginning with the FRW metric, the evolution of the scalar field $\phi$ is described by the Klein-Gordon equation \cite{kinney1997hamilton}
\begin{equation}\label{infl-eq-motion}
\ddot{\phi}+3 H \dot{\phi}+ V'(\phi) = 0,
\end{equation}
where the Hubble parameter $H$ is $H\equiv \dot{a}/a$.

Typically, one first specifies the scalar field potential $V(\phi)$ and then determines how $p$ and $\rho$ relate and evolve with time. Here, we will begin with the assumption of a Van der Waals equation of state and then determine the effective potential $V(\phi)$ that can produce an early-time accelerated expansion that matches with observation.

The Friedmann equations (eqs.~(\ref{Friedmann1}), (\ref{Friedmann2})) can be written in terms of the scalar field $\phi$ as:
\begin{eqnarray}
  \label{inflFRIEDMANN1}
H^2 = \left(\frac{\dot{a}}{a}\right)^2 &=& \frac{8 \pi G}{3} \left(V(\phi)+\frac{1}{2}
 \dot{\phi}^2\right) \\
\frac{\ddot{a}}{a}&=&\frac{8 \pi G}{3} \left(V(\phi)-
 \dot{\phi}^2\right).
  \label{inflFRIEDMANN2}
\end{eqnarray}
Inflation occurs when $\frac{\ddot{a}}{a} > 0$, which is equivalent to $\dot{\phi}^2<V(\phi)$. A de Sitter expansion occurs in the limit where $\dot{\phi}\rightarrow0$. 

Combining eqs.~(\ref{Friedmann1}), (\ref{rhodot}), (\ref{inflrhoP}), (\ref{inflP}), and the relationship
\begin{equation}
\label{eq:dphidt}
\left(\frac{d\phi}{d t}\right )^2=\rho+p,
\end{equation}
we find
\begin{equation}\label{drhodphi}
\frac{d\rho}{d\phi}=\pm\sqrt{24 \pi G \rho(\rho+p)}
\end{equation}
where the sign indicates whether the scalar field potential $V(\phi)$ is bounded from above (+) or below (-). Eqs.~(\ref{inflrhoP}) and (\ref{inflP}) may be rearranged to give
\begin{equation}\label{Vofphi}
V(\phi)=\frac{1}{2} (\rho-p).
\end{equation}

The inflationary slow-roll parameters are 
\begin{equation}
\label{eq:epsilonofphi}
\epsilon(\phi)=\frac{1}{16\pi G}\left (\frac{V'(\phi)}{V(\phi)}\right )^2
\end{equation}
and
\begin{equation}
\label{eq:etaofphi}
\eta(\phi)=\frac{1}{8\pi G}\frac{V''(\phi)}{V(\phi)}.
\end{equation}
It is convenient to define the amount of inflationary expansion in terms of the number of e-foldings ($N$) of the scale factor 
\begin{equation}
\label{eq:efoldsofphi}
N=2\sqrt{\pi G}\int_{\phi_e}^{\phi}\frac{d\phi'}{\sqrt{\epsilon(\phi')}},
\end{equation}
where $\phi_e$ is the field value at the end of inflation (where $\epsilon=1$).

Using eq.~(\ref{Vofphi}) to re-express eqs.~(\ref{eq:epsilonofphi}), (\ref{eq:etaofphi}), and (\ref{eq:efoldsofphi}) in terms of $\rho$, we find
\begin{equation}
\epsilon(\rho)=\frac{3}{2}\frac{\left [\sqrt{\rho \left (\rho +p\right )}\left (1-\frac{dp}{d\rho}\right )\right ]^2}{\left (\rho -p\right )^2}
\end{equation}
\begin{equation}
\eta(\rho)=\frac{3}{2}\frac{\left (1-\frac{dp}{d\rho}\right ) \left (\rho+p+\frac{dp}{d\rho}+1\right )}{\rho-p}
\end{equation}
\begin{equation}
N=\frac{1}{3}\int_{\rho_e}^{\rho}\frac{d\rho'}{\rho'+p(\rho')}.
\end{equation}
The above equations are in turn used to calculate the spectral index $n_s$ and the tensor-to-scalar ratio $r_{t/s}$.
\begin{eqnarray}
\label{n_s}
n_s &=& 1-6\epsilon+2\eta=0.9667\pm 0.0040 \\
\label{r_ts}
r_{t/s} &\approx& 16\epsilon\le 0.11
\end{eqnarray}
The numerical constraints given in eqs.~(\ref{n_s}) and (\ref{r_ts}) are evaluated for $N \gtrsim 50-60$ e-foldings, consistent with the Planck 2015 results.\cite{Ade:2015xua}

The constraints that we use to evaluate the three Van der Waals models are:
\begin{enumerate}
\item
\label{itm:realvalued}
$\frac{d\phi}{dt}$ is real-valued at all times
\item
\label{itm:epsilonbeginning}
$\epsilon\ll 1$ at the beginning of inflation
\item
\label{itm:etabeginning}
$\eta \ll 1$ at the beginning of inflation
\item
\label{itm:epsilonequals1}
$\epsilon=1$ at some point during the evolution of the scalar field $\phi$ (indicating the end of inflation)
\item
\label{itm:nvalue}
$n_s=0.9667\pm 0.0040$ at $\sim 50$ e-folds before the end of inflation
\item
\label{itm:rvalue}
$r_{t/s}\leq 0.11$ at $\sim 50$ e-folds before the end of inflation
\end{enumerate}

\section{One-parameter model}
\label{sec:oneparametermodel}
The reduced Van der Waals equation of state is given by eq.~(\ref{eq:1parameterEoS}). The evolution of the scale factor as a function of energy density is 

\begin{equation}
a(\rho)=C\left (\frac{3+8w-10\rho+3\rho^2}{\rho^2} \right )^{\frac{1}{2(3+8w)}}\left (\frac{3\rho-5-2\sqrt{4-6w}}{3\rho-5+2\sqrt{4-6w}} \right )^{\frac{2w-3}{3(3+8w)\sqrt{4-6w}}},
\end{equation}
where $C$ is an integration constant.\cite{jantsch2016van}  The initial value of $\rho$ (when $a\rightarrow 0$) is

\begin{equation}
\rho_0=\frac{5-2\sqrt{4-6w}}{3}.
\end{equation}

It should be noted that at $\rho_0$, $p=-\rho_0$ for $w=\frac{1}{8}$, indicating that this one-parameter model is capable of producing de Sitter expansion in the early universe.  When evaluated at the beginning of inflation ($\rho=\rho_0$), we find that $\epsilon(\rho_0)=0$, but $\eta(\rho_0)<-3$ for $0\le w\le \frac{2}{3}$ (which violates constraint \ref{itm:etabeginning}).  

\section{Two-parameter model}
\label{sec:twoparametermodel}
The Van der Waals equation of state parametrized in terms of the critical point is given by eq.~(\ref{eq:2parameterEoS}).  The scale factor as a function of $\rho$ is

\begin{equation}
a(\rho)=C\left [ \frac{\frac{3\gamma}{8\rho_c^2}\rho^2-\frac{27\gamma+1}{24\rho_c}\rho+\gamma+1}{\rho^2}\right]^{\frac{1}{6(\gamma+1)}}\left [\frac{\frac{3\gamma}{4\rho_c^2}\rho+\frac{27\gamma+1}{24\rho_c}-\sqrt{\left (\frac{27\gamma+1}{24} \right )^2-\frac{3\gamma}{2\rho_c^2}(\gamma+1)}}{\frac{3\gamma}{4\rho_c^2}\rho+\frac{27\gamma+1}{24\rho_c}+\sqrt{\left (\frac{27\gamma+1}{24} \right )^2-\frac{3\gamma}{2\rho_c^2}(\gamma+1)}} \right]^{\frac{\frac{27\gamma+1}{24\rho_c(\gamma+1)}+\frac{1}{3\rho_c}}{6\sqrt{\left ( \frac{27\gamma+1}{24\rho_c}-\frac{3\gamma}{2\rho_c^2}(\gamma+1)\right )}}}
\end{equation}
where $C$ is a constant of integration. If we restrict ourselves to real-valued solutions for $a(\rho)$ that monotonically decrease with increasing $\rho$, we may restrict the allowed range of the parameter $\gamma$ to
\begin{equation}
\label{eq:firstgammarestriction}
-\frac{24}{43}\le\gamma\le-\frac{8}{27}
\end{equation}
or
\begin{equation}
\label{eq:secondgammarestriction}
 0\le\gamma\le\frac{-72+32\sqrt{6}}{45}.
\end{equation}
For this model, the initial value of $\rho$ is
\begin{equation}
\rho_0=\frac{\frac{27\gamma+1}{24\rho_c}-\sqrt{\left ( \frac{27\gamma+1}{24\rho_c}\right )^2-\frac{3\gamma}{2\rho_c^2}(\gamma+1)}}{\frac{3\gamma}{4\rho_c^2}}.
\end{equation}

Numerically combining eqs.~(\ref{eq:firstgammarestriction}) and (\ref{eq:secondgammarestriction}), the requirement that $|\eta |\ll 1$ further constrains the allowed range of $\gamma$ to be
\begin{equation}
\label{eq:thirdgammarestriction}
0.105\le \gamma \le 0.142
\end{equation}
(with $0.105\le \gamma <0.121$ for $\eta<0$ and $0.121<\gamma \le 0.142$ for $\eta >0$).
For the allowed range of $\gamma$ (eq.~(\ref{eq:thirdgammarestriction})) in the two-parameter model, we find that constraint \ref{itm:realvalued} is violated. This is because immediately after inflationary expansion begins, $\rho+p<0$ implying that $d\phi/dt$ becomes complex-valued (from eq.~(\ref{eq:dphidt})).

\section{Three-parameter model}
\label{sec:threeparametermodel}

The scale factor for the three-parameter equation of state is
\begin{equation}
\label{eq:aofrho}
a(\rho)=C\left[\frac{\alpha\beta\rho^2-(\alpha+\beta)\rho+\gamma+1}{\rho^2}\right]^{\frac{1}{6(\gamma+1)}}\left[\frac{2\alpha\beta\rho+\alpha+\beta-\zeta}{2\alpha\beta\rho+\alpha+\beta+\zeta}\right]^{\frac{\frac{\alpha+\beta}{2(\gamma+1)}+\beta}{3\zeta}},
\end{equation}

where $C$ is an integration constant, $\zeta\equiv\sqrt{(\alpha+\beta)^2-4\alpha\beta(\gamma+1)}$, and $\zeta$ is real-valued\footnote{We demand that $\zeta$ be real valued to ensure that the scale factor $a$ remains real valued.}. Setting $a(\rho_0)=0$ yields
\begin{equation}
\label{eq:initialrho}
\rho_0=\frac{\alpha+\beta-\zeta}{2\alpha\beta}.
\end{equation}
It can be shown that at the beginning of inflation
\begin{equation}
\epsilon(\rho_0)=0,
\end{equation}
which automatically satisfies constraint \ref{itm:epsilonbeginning} ($\epsilon \ll 1$ at the beginning of inflation).  

Despite not being explicitly ruled out, we find the three-parameter Van der Waals equation of state to be highly fine-tuned. Setting  $\alpha=0.008$, $\beta=0.0002$, and $\gamma=-0.26$, we find  $\epsilon(\rho_0)=0$ and $\eta(\rho_0)=-0.01655$. In this model, the universe will undergo $N\simeq 56$ e-foldings of expansion and have a spectral index and tensor to scalar ratio of $n_s=0.9669$ and $r_{t/s}\simeq 0$, respectively. This model is highly fine-tuned in that a variation in any of the Van der Waals parameters ($\alpha$, $\beta$, or $\gamma$) by as little $\sim 1\%$ causes the three parameter Van der Waals model to violate at least one of the constraints of section \ref{sec:InflationwithVdWFluid}.

\begin{figure}
\center
\includegraphics[width=0.7\textwidth]{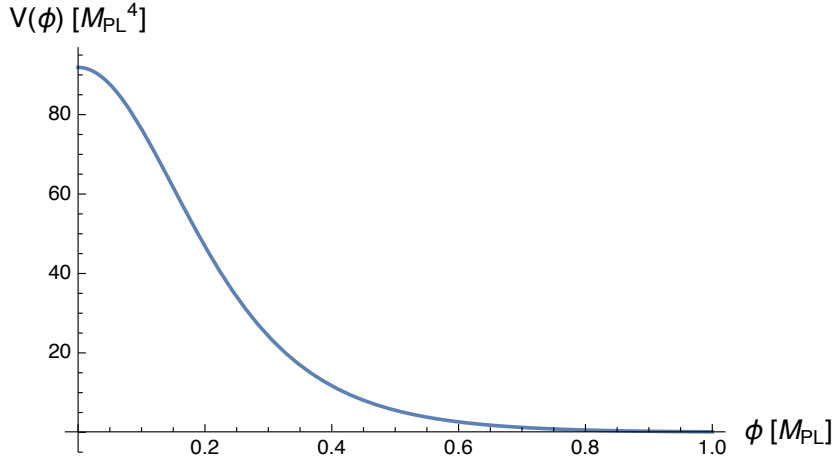}
\caption{The scalar field potential $V(\phi)$ for a three-parameter Van der Waals equation of state.  The parameter values used are $\alpha=0.008$, $\beta=0.0002$, and $\gamma=-0.26$.}
\label{fig:VofPhi}
\end{figure}

Using eqs.~(\ref{drhodphi}), (\ref{Vofphi}), and the parameter values of $\alpha$, $\beta$, and $\gamma$ mentioned above, we may reconstruct the inflaton potential. In this model, we find that the Van der Waals equation of state behaves as a scalar field on a hilltop potential. A numerical reconstruction of the scalar field potential $V(\phi)$ is shown in figure \ref{fig:VofPhi}.  From our reconstruction, we find that the characteristic scale of the inflaton potential is of the same order of magnitude as the Planck scale. If we require that all inflationary physics occur below the Planck scale, we may rule out this three parameter model as well.

\section{Conclusions}\label{sec:Conclusions}

In this work, we investigated the viability of three different parametrizations of a Van der Waals fluid as a model of inflation in the early universe. The models each contained one, two, or three free parameters in the Van der Waals equation of state. We found that the three Van der Waals models considered may produce an epoch of de Sitter-like expansion in the early universe (when the scale factor $a\rightarrow 0$), but two of the three models considered violate observational constraints and the third model is highly fine-tuned.

The first model considered (with only one free parameter) predicts a non-slow rolling field solution (specifically the slow roll requirement $|\eta|\ll 1$ is violated) at the beginning of inflation. Demanding a slow-roll inflationary solution (e.g. the slow roll parameters $\epsilon$ and $\eta$ are small) allows us to numerically constrain the constant $\gamma$ in the two-parameter Van der Waals equation of state. This two-parameter Van der Waals equation of state may be ruled out when we demand that $d\phi/dt$ remain real valued.

We did find that there exists a viable but highly constrained region parameter space for the three parameter Van der Waals equation of state that does not violate any of the six constraints listed in section \ref{sec:InflationwithVdWFluid}. These values center around $\alpha=0.008$, $\beta=0.0002$, and $\gamma=-0.26$. These values are rather fine-tuned in that a variation of any of these parameters by as much as $\sim 1\%$ may be excluded. Nonetheless, we can reconstruct an effective hilltop-like inflationary potential from this three parameter Van der Waals equation of state. This model may be ruled out, however, if we require a seventh constraint: that the inflaton potential does not surpass the Planck scale.

\bibliographystyle{unsrtnat}
\bibliography{bibliography}

\end{document}